EXPLORING ATTACK RESILIENCE IN DISTRIBUTED PLATOON CONTROLLERS WITH
MODEL PREDICTIVE CONTROL

By

TASHFIQUE HASNINE CHOUDHURY

A THESIS PRESENTED TO THE GRADUATE SCHOOL
OF THE UNIVERSITY OF FLORIDA IN PARTIAL FULFILLMENT
OF THE REQUIREMENTS FOR THE DEGREE OF
MASTER OF SCIENCE

UNIVERSITY OF FLORIDA

2023

I dedicate this thesis to my loved ones, who have been my source of strength throughout this academic journey, for their constant support and encouragement. Thank you to my family and friends for always being there for me and for your constant motivation and inspiration. Your presence has made this difficult journey even more gratifying. I would want to dedicate this thesis to my mentors, advisors, and field researchers, who have been an ongoing fountain of inspiration to me. Their unflinching dedication, game-changing contributions, and relentless pursuit of excellence inspire me to strive for greatness. This work is a testimonial to the collaborative efforts of everyone who has helped me grow, and I am grateful and proud to have you all in my life.

ACKNOWLEDGEMENTS

I want to take this opportunity to convey my heartfelt gratitude to my thesis advisor, Dr. Sandip Ray, for his undying belief in me and unparalleled cooperation in helping me defend my master's thesis and research. I thank him for steering me in the proper direction, keeping track of my work and writing, and boosting and inspiring me along my journey. I shall be eternally grateful to have had such an advisor and mentor to lead me through my master's thesis and degree.

I would want to thank Dr. Swarup Bhunia and Dr. Zoleikha Biron for agreeing to take time out of their busy schedules to be a part of my wonderful academic experience and support me all throughout. I would also want to thank my former and current laboratory associates at the Warren B. Nelms IoT Institute, Srivalli Boddupalli, and Chengwei Duan, for their help, advice, and thoughtful remarks on my research.

Finally, I want to express my appreciation to my family for believing in me and investing their hard-earned money in me. None of this would have been possible without their persistent support and assistance. They are the reason I get up every day, flourish, and strive to improve day by day. I would like to thank my mother, Dr. Gulshan E Jahan, in particular, for being my fortress and always being there for me when I needed her during my times of struggle. I conclude by praying to God and expressing love and gratitude to everyone who has helped me get to where I am.

# TABLE OF CONTENTS

LIST OF TABLES

LIST OF FIGURES

LIST OF OBJECTS

.

Abstract of Thesis Presented to the Graduate School
of the University of Florida in Partial Fulfillment of the
Requirements for the Degree of Master of Science

EXPLORING ATTACK RESILIENCE IN DISTRIBUTED PLATOON CONTROLLERS WITH MODEL PREDICTIVE CONTROL

By

Tashfique Hasnine Choudhury

August 2023

Chair: Sandip Ray
Major: Electrical and Computer Engineering

The extensive use of distributed vehicle platoon controllers has resulted in several benefits for transportation systems, such as increased traffic flow, fuel efficiency, and decreased pollution. The rising reliance on interconnected systems and communication networks, on the other hand, exposes these controllers to potential cyber-attacks, which may compromise their safety and functionality. This thesis aims to improve the security of distributed vehicle platoon controllers by investigating attack scenarios and assessing their influence on system performance. Various attack techniques, including man-in-the-middle (MITM) and false data injection (FDI), are simulated using Model Predictive Control (MPC) controller to identify vulnerabilities and weaknesses of the platoon controller. Countermeasures are offered and tested, that includes attack analysis and reinforced communication protocols using Machine Learning techniques for detection. The findings emphasize the significance of integrating security issues into their design and implementation, which helps to construct safe and resilient distributed platoon controllers.

CHAPTER 1
INTRODUCTION

The rapid adoption of connected autonomous vehicles (CAVs) is undergoing a substantial transition in the transportation sector globally. These cars include a wide range of technologies that let them do difficult maneuvers without the help of a person, from simple driver assistance tools to the more sophisticated, fully automated driving.

Their ability to link and communicate is one of the most crucial features of the emerging field of autonomous cars. This implies that they are capable of seamlessly interacting and exchanging data not only with other vehicles, which is known as Vehicle-to-Vehicle (V2V) communication, but also with the local infrastructure, such as traffic lights, toll booths, or parking systems, which is known as Vehicle-to-Infrastructure (V2I) communication. These autonomous vehicles may also communicate with numerous V2IoT (Variable to Internet of Things) devices, such as cell phones, home automation systems, or other electronic gadgets. This level of connectivity, which is generally referred to as V2X or "vehicle-to-everything," has quickly established itself as a pillar of the ecosystem for autonomous vehicles. V2X communication is vital to improving and refining the functioning of autonomous vehicles because it enables cooperative and dynamic information sharing between vehicles and their surroundings.

Platooning is one of the most commonly used connected driving applications. For a group of CAVs, this application leverages V2X communication technology to enable effective vehicle-following patterns. In this platoon, a leader CAV sets the path, and the other vehicles closely follow while keeping a very tiny yet secure spacing between one another. This strategy involves more than just following the leader. Depending on how platooning is implemented, the following vehicles may potentially communicate with one another. This direct interactive communication improves the platoon's stability, making autonomous driving a lot smoother and more secure.

The effects of platooning go beyond security. Additionally, the utilization of road infrastructure and fuel economy have significantly increased as a result of this strategy. Vehicles can decrease wind resistance which results in reduced fuel consumption. Additionally, making the most use of the already-existing infrastructure by keeping tight distances and coordinating movement, effective use of road space may aid in reducing traffic congestion and enhancing overall traffic flow. CAVs and related technologies are thereby altering how we view and use transportation networks, offering a more effective, secure, and environmentally friendly, healthy future.

However, due to intrinsic flaws in perception technology, which includes sensors and V2X communication systems, CAV platooning is vulnerable to serious security vulnerabilities. The problem with perception attacks is that they target the very channels that give CAVs an understanding of their surroundings. The effects of a successful attack on a CAV's perception are extensive and may go beyond the specific vehicle, wreaking havoc across the entire platoon and possibly even impacting the local transportation network. These attacks can be quite simple to carry out yet extremely effective, which is alarming. For instance, a malicious actor could alter perceptual inputs to trick the CAV. Through the use of V2X, a "Reduce Speed" command given by the lead vehicle might be tampered with and altered into a "Increase Speed" command, resulting in catastrophic collisions with other vehicles in the platoon. Another possibility is that the attacker might tamper with the sensors that collect vital data, like the distance between CAVs. This might lead to errors in judgment and even start a series of crashes within the platoon. In a worst-case scenario, an attacker might launch a Denial of Service (DoS) attack and stop the platoon members from communicating over V2X, which, as a result, nullifies the basic connectivity aspect of a CAV. The disruption might significantly lower the platoon's operational effectiveness and safety. In addition, a determined enemy with sufficient resources might simultaneously target several sensory channels on one or perhaps several CAVs in the platoon. This creates a wide range of potential attacks, which increases the necessity for effective defense measures to counter a wide range of attack capabilities. Therefore, it is crucial to develop trustworthy, real-time resiliency solutions for platooning systems in such hazardous driving settings. Taking into account the realities and eccentricities of actual application settings, these systems ought to be able to fight against a variety of perception attacks, and this broad attack spectrum must be investigated and analyzed in order to accomplish that.

In this thesis, we conducted a comprehensive analysis of possible attack scenarios, their impacts, and the potential mechanisms for detecting such attacks. Furthermore, we also put forth a set of proposed solutions for fortifying the development of resiliency of such a platoon controller that operates using the MPC method. To comprehend the potential effects of attacks on these platoon controllers, it is essential to examine their nature. By examining these impacts, we can streamline the process of creating robust solutions to counteract them.

The extensive range of possible attack scenarios includes various forms of attacks that target single or multi-vehicle V2V communication channels. Some of these attacks are exceptionally aggressive and more effective than others. These attacks could jeopardize the safety of the vehicle platoon, hinder its efficiency, or even induce discomfort. In each case, it's essential to thoroughly analyze and visually represent these attacks and their potential impacts. The primary focus of this thesis lies in implementing and investigating such attack orchestration techniques and portraying their impacts on the platoon controller.

This specific platoon controller stands apart from other platoon controllers due to its unique method of operation. The controller makes multiple calculations through several iterations, leading to a distinct pattern of impacts when compared to those controllers. This uniqueness presents a challenge in creating an organized and effective detection mechanism. To tackle this challenge, we have developed a two-phase detection mechanism that uses an Extreme Learning Machine (ELM) model. This detection mechanism is tailored to the unique requirements of these platoon controllers and has proven successful in identifying potential attacks launched on the controller's communication channels. Through this system, we can significantly improve the security of the communication channels and ensure the robust operation of the platoon controller, effectively mitigating potential attacks.

Chapter 2 lays the framework for the research by providing a thorough explanation of the decentralized platoon controller and the corresponding work in the area of platoon security. This section provides a brief overview of platooning's fundamental ideas, actual implementation, and communication topology.

Chapter 3 delves deeply into the discussion of the threat model and categorizations of potential attacks. This section of the work is devoted to comprehending and deconstructing the complexity of the selected threat model, providing a thorough examination of its different aspects and complexities. The study can mimic real-world events, identify potential risks and dangers, and develop ways to minimize these threats by employing a specialized adversary model for all experiments and analyses.

Chapter 4 highlights the architecture that explores potential threats in the context of a decentralized platoon controller. It discusses not only the restrictions and shortcomings of this system but also places a strong emphasis on the evaluation of attack impacts. It explores the potential effects and ramifications that attacks may have on the overall effectiveness, safety, and efficiency of the system.

Chapter 5 discusses a detection method for these attacks, focusing on how thoroughly it has been assessed in terms of detection precision and experimental analysis. The chapter also provides a thorough analysis of the algorithms and models used for the task.

Chapter 6 focuses on drawing conclusions and work that can be done in future endeavors. To have a thorough understanding of the topic, the key findings and revelations discovered in the earlier chapters are summarized and examined in this section. The chapter also gives an idea of a proactive strategy for tackling anticipated difficulties and uncertainties that might appear in future undertakings by putting out a strong and resilient response.

# CHAPTER 2
# BACKGROUND AND RELATED WORK

## 2.1 Background

### 2.1.1 Fundamental Concepts of Platoons

Vehicle-to-Vehicle (V2V) communication and on-board sensors allow a group of cars to act together as a "platoon" in vehicular control. They exchange real-time data like position, velocity, and acceleration. Platooning's main objective is to increase fuel efficiency while maintaining safe separations between each vehicle, or "headway". Without V2V communication, this is especially crucial because bigger gaps are required to provide sensors enough time to react to potential threats or changing environmental conditions. For CAVs, platooning applications are often developed as either centralized or distributed control systems, each with a different information flow topology. In a centralized system, the "leader", which is often the first vehicle in the platoon, controls driving decisions for each vehicle inside the platoon. The following vehicles' speed, direction, and general functioning are dictated by this vehicle, which also establishes the driving guidelines for the entire platoon. As opposed to this, with a distributed platoon system, each vehicle has a control system that governs how it modifies its speed. Each vehicle can make decisions using its particular circumstances and the shared information it receives, thanks to this decentralized approach. In such a system, a platoon controller can make decisions using information from any subset of the platoon's vehicles. This MPC-based decentralized strategy offers a certain amount of adaptability and individual autonomy while yet upholding the platoon system's cohesion and mutual advantages.

### 2.1.2 MPC-based Decentralized Platoon Controller Fundamentals

A MPC-based decentralized platoon controller [10] would use a separate controller for each vehicle in the platoon. MPC is a form of a control algorithm that predicts the system's future behavior using a model of the system. MPC determines the optimal control actions by solving an optimization problem at each time step. The issue is written in such a way that it minimizes a cost function over a prediction horizon while keeping control actions and system behavior limitations in mind. Each vehicle's controller would utilize a model of the vehicle and its immediate surroundings to forecast future platoon behavior and decide the best control actions for the vehicle. The usage of MPC in this scenario provides a number of advantages. It can handle multi-input, multi-output systems as well as behavioral limitations, which are frequent in vehicle platooning. It may also optimize over a time horizon, which can aid in anticipating and avoiding potential hazards. Each vehicle in the platoon would communicate with its neighbors to share information about their current conditions and future plans. This would allow each vehicle's controller to consider its neighbors' conduct while determining its own control. In this thesis, we chose a controller in which the platoon requires the information from the leading vehicle to direct all subsequent vehicle motion, but the leading vehicle is not regulated by the platoon controller and can freely decide its own motion. For this controller, it is assumed that the platoon is a pure CAV platoon with a leading vehicle and many following vehicles and that the true implementation has the information flow topology depicted in Figure 2-1.

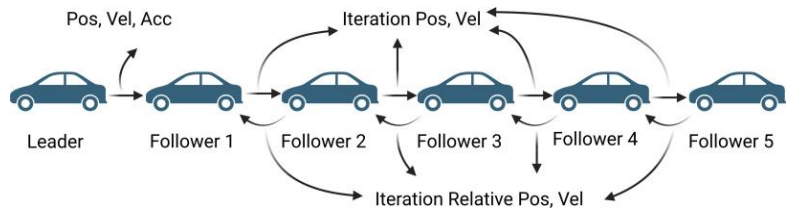

Figure 2-1. Information Flow Topology of the Target Platoon Controller

### 2.1.3  Operational Methodologies and Computational Procedures

The vehicle platooning controller, which we will use in this thesis, is a control system that optimizes vehicle platooning via online learning and distributed optimization. This platoon follows the procedure indicated in Figure 2-1. The rate of acceleration or deceleration of the vehicle at each interval will be the deciding factor in this MPC-based closed-loop system. MPC's control technique makes use of the vehicle's past states, which are fed into a predictive model. These predicted outcomes are then given to the optimization module. The optimizer approaches platoon control as a convex optimization problem with constraints ensuring a safe distance, speed limit, and acceleration boundaries. The optimizer's output is subsequently used as the vehicle's actuation input. A representation of this sequence is provided in Figure 2-2.

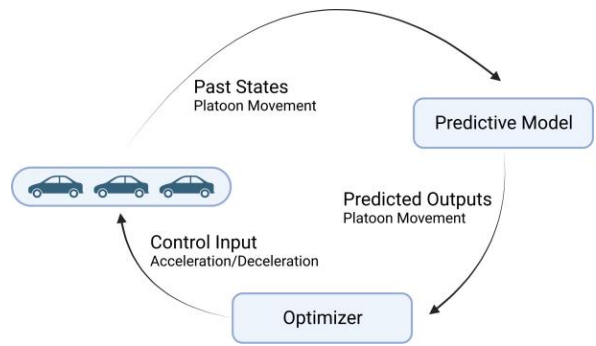

Figure 2-2. Working Flow of the MPC-Based Platoon Controller

Equations 2-1 and 2-2 describe the discrete-time longitudinal dynamics of the prediction model in Figure 2. Table 2-1 contains a glossary of the words used in the calculations above. The range $[k\tau, (k+1)\tau]$ determines what a control step is. Additionally, each vehicle completes iteration stages within a control step in order to optimize the problem.

$$v_i(k+1) = v_i(k) + u_i(k)\tau \qquad (2\text{-}1)$$

$$x_i(k + 1) = x_i(k) + v_i(k)\tau + \frac{u_i(k)}{2}\tau^2 \qquad (2\text{-}2)$$

Then the predicted values from the predictor is passed on to the optimizer, which minimizes a strictly convex cost function which is shown in equation 2-3. Penalty is added on all related distances and all related velocities between each adjacent vehicle, and the absolute value of the acceleration itself is the cost that needs to be minimized. The optimization is carried out with the restrictions of acceleration limit, velocity limit, and safety distance. In order to maintain a consistent time headway, the safety distance is computed in an adaptive manner with respect to velocity. Equations 2-4, 2-5, and 2-6 display the constraints. The equations 2-7 and 2-8 are used to represent the control dynamics. The optimal control value is obtained by Newton's method of optimization using the control equations.

$$Min \Gamma(u) = \sum_{i=1}^{n} \frac{1}{2} z_i^T(k+1) Q_\alpha z_i(k+1) + z_i(k+1)^{'T} Q_\beta z_i(k+1)^{'} + \frac{\tau^2}{2} u_i^T(k) u_i(k)] \qquad (2\text{-}3)$$

$$a_{min} \leq u_i(k) \leq a_{max} \qquad (2\text{-}4)$$

$$v_{min} \leq v_i(k+1) \leq v_{max} \qquad (2\text{-}5)$$

$$x_{i-1}(k+1) - x_i(k+1) \geq (L + p\tau v_i(k+1)) \qquad (2\text{-}6)$$

$$z_i(k + 1) = x_{i-1}(k + 1) - x_i(k + 1) - (L + p\tau v_i(k + 1) + \delta) \tag{2-7}$$

$$z_i'(k + 1) = v_{i-1}(k + 1) - v_i(k + 1) \tag{2-8}$$

Table 2-1. Glossary of Notations

| Term | Definition |
|---|---|
| $u_i(k)$ | The acceleration of follower vehicle $i$ at time step $k$ |
| $v_i(k)$ | The velocity of follower vehicle $i$ at time step $k$ |
| $x_i(k)$ | The position of follower vehicle $i$ at time step $k$ |
| $\tau$ | The sample length in seconds |
| $k$ | The control step. $k = 1, 2, 3...$ |
| $z_i(k)$ | The spacing between follower vehicle $i-1$ and $i$ for next time step $k$ |
| $z_i'(k)$ | The relative speed fluctuation of follower vehicle $i-1$ and $i$ for next time step $k$ |
| $Q_\alpha, Q_\beta$ | Penalty matrix |
| $L$ | Length of vehicle |
| $n$ | Total number of followers. For our implementation, n = 6 |
| $p$ | Adaptive spacing constant |
| $\delta$ | Auxiliary constant to avoid in-feasibility |
| $a_{min}, a_{max}$ | Minimum and maximum acceleration |
| $v_{min}, v_{max}$ | Minimum and maximum velocity |
| $u\_ite(t)$ | The iterative acceleration value of a follower vehicle at time $t$ |
| $v\_ite(t)$ | The iterative velocity value of a follower vehicle at time $t$ |
| $x\_ite(t)$ | The iterative position value of a follower vehicle at time $t$ |
| $zv\_ite(t)$ | The iterative relative velocity value of a follower vehicle at time $t$ |
| $zx\_ite(t)$ | The iterative relative position value of a follower vehicle at time $t$ |
| $fvX$ | The following vehicle number X |

This particular platoon implementation uses a double loop method based distributed algorithm to obtain the final value of computed acceleration. The loops are called the Primal loop and the Dual loop. During one control step $[k\tau, (k + 1)\tau]$, the following vehicle's acceleration is kept constant $u(k\tau)$. The iteration steps during the control step will only determine the acceleration for the next control step, which is $u((k + 1)\tau)$. The primal loop will stop if the inequality $u\_ite(i+1) - u\_ite(i) \leq 0.01$ is satisfied. On the other hand dual loop will stop if all gaps between each pair of adjacent vehicles are larger than the safety gap. However, this implementation will have no idea how many primal and/or dual loop iterations will be executed prior to execution, and so a maximum number of iteration limit is set in order to stop the loops even if the stop conditions are not met. The flow diagram of this implementation is shown in Figure 2-3

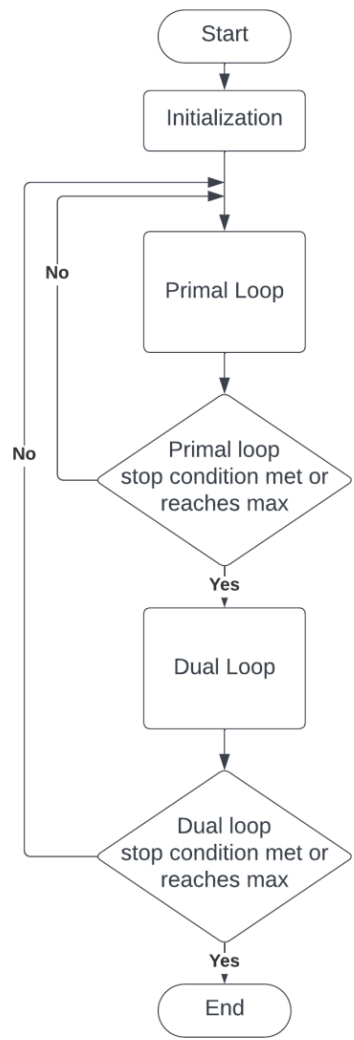

Figure 2-3. Flow Chart for the Primal and Dual Algorithm

## 2.2 Related Work

The development of defenses against attacks on the perception channels inside vehicular platoons has been the subject of numerous studies. Boddupalli *et al.*[5] have portrayed an overview of CAV resilience approach based on machine learning, based on which our approach draws inspiration from. While V2V perception attack impacts had been demonstrated in the previous work by Boddupalli *et al.*[4][7], the V2I perception channel attack impacts had been highlighted by Mendoza *et al.*[16]. To comprehend vehicle dynamics in a platoon, for instance, Biroon *et al.*[3] developed a partial differential equation model. They then utilized a similar observer to detect false vehicle injection attacks. Similar to this, Mokari *et al.*[17] developed a resilient control system combining switching system theory, a high-gain observer, and a pseudo-Laplacian matrix to detect false input. To recognize and isolate FDI assaults within a platoon controlled by a cloud-based control center, Huang *et al.*[11] additionally developed a resilient state observer coupled with an adaptive detection threshold approach.

Furthermore, a significant percentage of this field's research is focused on learning algorithm-based approaches. Approaches using Machine Learning in CAV security had been briefly overviewed by Boddupalli *et al.*[6]. While Zhao *et al.*[23] built a cloud-based sandboxing framework for detecting fake velocity and location injections, Cabelin *et al.*[8] constructed an SVM-based intrusion detection system for each vehicle in a platoon. Khanapuri *et al.*[14] proposed an FDI attack detection and mitigation technique using unaltered local sensors, while Alobiti *et al.*[1] provided a real-time anomaly detection system for leader acceleration channel attacks.

The use of a sliding mode observer [12] [2] [13] [22] is another popular technique for identifying and assessing fake data injection attacks in vehicle platoons. State estimators are also often used, and Zuo *et al.*[25] have introduced a security approach using distributed state estimators. Dutta *et al.*[9] and associates also created a novel robust distributed state estimator to reduce state estimation mistakes during an attack. Both Lou *et al.*[21] and Song *et al.*[19] promoted secure state estimation for platoon-level inter-vehicle network attacks. A technique to recognize and lessen the negative impacts of false acceleration data injection assaults in a CACC-based platoon was also created by Micheal *et al.*[20]. Based on their designed BLF-PPC controller, Zhou *et al.*[24] presented an attack diagnosis and compensation scheme. For platoons with uncompromised leader channels, Petrillo *et al.*[18] offered a secure adaptive control system, and Lu *et al.*[15] created a robust sensor fusion technique.

However, many of the current methods, meanwhile, are meant to address a particular kind of FDI attack and might not be able to successfully fend off single- or multi-channel attacks on arbitrary arrangements of perception channels. Additionally, a lot of these systems just take into account one particular type of misleading data and do not take other types into account. As a result, these methods could not be flexible to a variety of implementations of perceptual opponents and controllers.

# CHAPTER 3
# THREAT MODEL AND ATTACK TAXONOMY

## 3.1 Identifying and Outlining the Threat Model

This thesis proposes a threat model that takes into account potential adversary behavior that could interfere with V2V communications. A "platoon" of vehicles, or a collection of them, constitutes a potential threat; more particularly, any vehicle not functioning as the leader vehicle. In other words, one or more of the vehicles in this formation may be the targets of tampered V2V messages sent by either the vehicles in front of them or the ones behind them.

In the context of the platoon, we use the term "ego vehicle" to denote the primary target vehicle. For the functioning of this platoon, essential parameters like preceding vehicle positions and velocities, along with relative positions and velocities of the following vehicles, are communicated to the ego vehicle through V2V during iterative steps. When these communication channels are compromised, the values for these parameters could become incorrect, misleading, or simply unavailable due to adversarial corruption.

Although there is a chance that follower/preceding vehicle communication could be compromised, this model assumes that there is some level of trust among the platoon. Particularly, it is presumed that the V2V connection between the leading vehicle and the first follower is reliable. In addition, each vehicle in the platoon is believed to have reliable on-board sensors, decision-making units, and actuarial modules. The threat model also includes scenarios where the MPC-based controller is subjected to single or multiple-channel adversarial attacks. This might cause a scenario in which all V2V communication channels among the follower vehicles are simultaneously compromised, leaving the ego vehicle without any trustworthy data gained through V2V communication.

The proposed process for exploring such attacks remains impartial to the source of the threat. A compromised communication network element or a rogue vehicle in the platoon that comes before or after the ego vehicle could be the source of the attack, *e.g.*, intentionally tampering with the communication protocol, the attacker could prevent message delivery or inject false data, further aggravating the situation.

## 3.2 Classification of Attacks: Attack Taxonomy

The V2V attack landscape is complex and diversified, including different communication attack techniques like MITM, masquerade, wormhole, and DoS through jamming or flooding. The possible attack surface for vehicular communications is continually changing since CAV applications and automotive security are still relatively new. There are frequently found newly created zero-day vulnerabilities and assaults. It is impractical to keep changing a resilience architecture in response to these fresh attacks. As a result, a successful resilience solution should not only defend against known threats but also display robustness against potential, undiscovered attacks in the future.

This approach, which was inspired by Boddupalli *et al.* [7], focuses on solving the problem by creating a thorough taxonomy of adversaries in CAV applications, shown in Figure 3-1. The crucial realization is that any V2V assault may be sufficiently described by a few well-defined features, despite the fact that attack tactics may change over time due to MITM attacks, compromised infrastructure components, or channel jamming.

The number of possible attack instances on a CAV application can quickly become unmanageable. Consider a simple adversary corrupting a single V2X communication channel of a CAV. The potential attacks in this scenario include (1) information mutation, (2) fabrication of fake information, or (3) prevention of information delivery through a jamming attack. The corruption can occur at different attack frequencies, and the extent of mutation can vary in form and magnitude. Evaluating the resilience of a system must account for this vast and complex attack space, which is challenging to achieve without the proper study and analysis of this attack space.

The attack taxonomy used enforces accurate characterization of the attack space in order to expedite the analysis of attacks. It makes use of the following six basic categorization criteria: (*i*) method of corruption, (*ii*) impact on the platoon , (*iii*) attack target, (*iv*) attack channels, (*v*) attack frequency, and (*vi*) attack mutation type. This taxonomy allows us to efficiently navigate the broad and complex universe of potential attacks, which would be challenging to do using only manual detection and resilience measures.

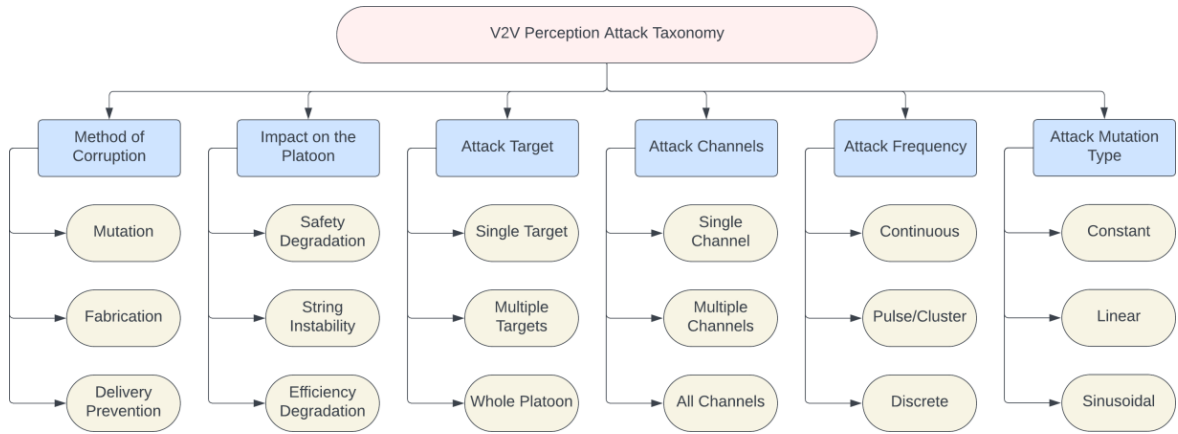

Figure 3-1. Taxonomy of Attack Space

# CHAPTER 4
# COMPREHENSIVE APPROACH TO ATTACK EXPLORATION

## 4.1 Process Outline, Bias Vector Formulation, and Analysis

### 4.1.1 Intricacies of Attack Strategies

We must implement a method of attack that has a wide attack surface. This suggests that our programming should be made to be scalable for attacks, allowing it the flexibility to handle threats of different intensities.

- One feature of this strategy is the development of a technique for producing bias vectors in a scalable manner. Our framework will be able to manage a variety of threat intensities and complexity by scaling up or down as necessary owing to this versatility.
- A further component of our methodology should be the creation of bias vectors that cover a wide range of possible attack scenarios. We may increase our resilience and preparedness against various potential threats by creating a system that covers a wide range of conceivable attacks.

There is some uncertainty about how many iterative steps each subsequent vehicle will need to reach its final commanded acceleration for each control step. For example, if we were to send the instruction, "Begin an iterative position attack for any following vehicle starting from the $10^{th}$ iteration step in the $20^{th}$ control step and ending at the $15^{th}$ iteration step in the $20^{th}$," this attack might not be viable. This is because a follower vehicle may only require 8 iteration steps in the $20^{th}$ step to acquire the final requested acceleration value for this control step. If this were the case, the vehicle would escape the loop early and advance to the next control step, rendering the attacker's efforts useless.

It is reasonable to assume that an attacker may carry out a series of consecutive attacks, each of which is introduced independently in various control phases. This is based on the fact that iterative communication occurs only at the start of each control step, which is quite short. As a result, it's unlikely that an attacker can sustain a continuous attack across numerous control stages. The scenario typically plays out within the confines of a fixed control step window, an assumption that allows for the potential compromising of any channel anytime an attack is initiated.

Understanding the order of communication inside each iterative stage is also critical. All succeeding cars first broadcast their 'x_ite' and 'v_ite' values to the vehicles that follow in each occurrence. The following cars also send their computed 'zx_ite' and 'zv_ite' values to the vehicles that preceded. The leader vehicle in the platoon is an exception to this pattern, as it does not receive information from its follower. This information flow implementation is shown in figure 4-1.

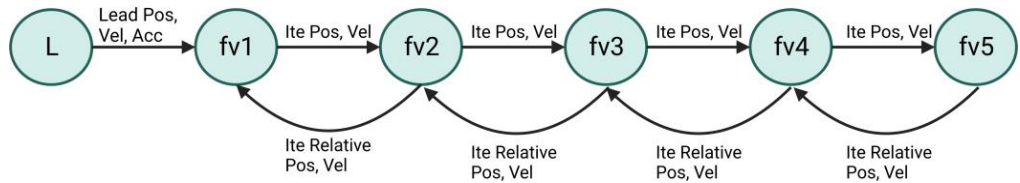

Figure 4-1. Information Flow Topology

Therefore, any attack would have to be sophisticated and well-planned in order to cause significant disruptions in the process, and thus, the creation of a complete bias vector for each control step is necessary. The bias vectors have to have bias values available for all the unknown number of iterations that executes within a control step. This comprehensive bias vector would be sufficient to manipulate the values of target channels, allowing for a more effective attack implementation.

### 4.1.2 Platform Specifications

The system we have developed for creating the bias vector allows for a considerable level of flexibility in investigating potential attacks. In the platoon model we employed, a maximum of 300 iterations is utilized within a control step. Hence, the bias vector must consist of at least 300 rows to ensure a successful attack. When generating a bias, our system considers several crucial factors:

1. ***Designated Following Vehicle:*** As part of our attempt to analyze the impact of attacks, we apply the same attacks to different following vehicles. The goal here is to gauge the differences in the impacts of these attacks. Essentially, this strategy involves generating a list of compromised vehicles, *e.g.*, iter_victim_list of $[1,3,5]$ means $fv1$, $fv3$, $fv5$ are attacked.

2. ***Attack Duration:*** This refers to the specific period of the control step during which an attack occurs. We can use this to simulate various kinds of attacks, including Cluster and Discrete attacks. We postulate that each following vehicle may experience one or more attack periods. We can also independently establish the duration of each attack period, *e.g.*, considering above mentioned victims, control_attackperiod_list can be $[[[10,20],[15,25]],[[20,30]],[[40,60]]]$, which means $fv1$ is attacked in all steps between the $10^{th}$ and $20^{th}$ control step, and between the $15^{th}$ and $25^{th}$ control step; $fv3$ is attacked in all steps between $20^{th}$ and $30^{th}$ control step; $fv5$ is attacked in all steps between the $40^{th}$ and $60^{th}$ control step.

3. ***Type of Compromised Channel:*** We can execute attacks on different channels within a given attack period. In the context of this controller, these channels might include 'x_ite', 'v_ite', 'zx_ite', and 'zv_ite', or any random combinations of these, *e.g.*, considering above mentioned victims and period, iter-malichannel-list can be $[[['x\text{-}ite','v\text{-}ite'],['v\text{-}ite']],[['zx\text{-}ite']],[['zv\text{-}ite']]]$, which means for $fv1$, its 'x_ite', 'v_ite' channels are attacked during the attack interval $[10,20]$ of the control step, and its 'v_ite' channel is attacked during the attack interval $[15,25]$. For $fv3$, its 'zx_ite' channel is attacked during the interval $[20, 30]$. For $fv5$, its 'zv_ite' channel is attacked during the interval $[40, 60]$ of the control step.

4. **Attack Frequency Type:** The attacker can specify the type of attack frequency. Options include cluster, discrete, or continuous frequencies. The attacker can freely impose multiple frequency attacks on the same target vehicle within the same attack period, *e.g.*, considering above mentioned cases iter_freq_type_ list can be $[[['Continuous','Continuous'], ['Cluster']], [['Continuous']], [['Cluster']]]$. Here, it means for $fv1$, the attack frequency type is $'Continuous'$ for both malicious channels during attack period $[10, 20]$, and the attack frequency type is $'Cluster'$ for 'v_ite' in $[15, 25]$. For $fv3$, the attack frequency type is $'Continuous'$ during the interval $[20, 30]$ of the control step. For $fv5$, the attack frequency type is $'Cluster'$ during the interval $[40, 60]$. We will not define the frequency type $'Discrete'$, and consider it as a special case of $'Cluster'$.

5. **Frequency Window:** If 'Cluster' attacks are intended, the attacker has the liberty to set distinct on/off window periods. The concept of a frequency window applies specifically to the 'Cluster' attack frequency type. However, for 'Continuous' attack type, we typically consider the window period as zero, *e.g.*, considering above stated cases iter_freqparavalue_ list may be $[[[[0], [0]], [[2, 8]]], [[[0]]], [[[1, 10]]]]$. If the frequency type is 'Continuous', the item in this list would be $'[0]'$. If the frequency type is 'Cluster', one example of the item is $'[1, 10]'$, which means that when we start the attack, the bias would be added for the first time step, then it will not be added for next 10 time steps. This process is executed periodically.

6. **Bias Attributes:** The attacker can launch attacks using different types of biases such as 'Constant', 'Linear', 'Sinusoidal', and so on. The initial value and parameters for different types of biases can be adjusted as desired, *e.g.*, keeping the above explained cases in mind, there will be two lists. First, iter_biastype_list as $[[['Constant','Constant'], ['Constant']], [['Linear']], [['Sinusoidal']]]$, which means for $fv1$, the bias type for both 'x_ite', 'v_ite' channels are $'Constant'$ during attack period $[10, 20]$ of the control step. For $fv3$, the bias type is $'Linear'$ for 'zx_ite' during the $20^{th}$ and $30^{th}$ control step.

For $fv5$, the bias type is $'Sinusoidal'$ for 'zv ite' during the interval $[40, 60]$. Second, iter_biasparavalue_list will be $[[[[3],[2]],[[4]]],[[[2,5]]],[[[10,0.5,0,5]]]]$ where:

- If the iter_biastype is 'Constant', it indicates a constant value (c), so the format of iter_biasparavalue would be $[c]$
- If the iter_biastype is 'Linear', it means there is a slope (m) and a constant (c), so the format would be $[m,c]$
- If the iter_biastype is 'Sinusoidal', it indicates there is an amplitude (A), frequency (f), theta ($\vartheta$) and a constant shift (c), so the format would be $[A, f, \vartheta, c]$

Ultimately, the comprehensive compilation of our iteration attack case list, intended to serve as the input for our bias generator as depicted in figure 4-2, is as follows:

```
iter_attackcase   =   [
                        iter_victim_list,
                        control_attackperiod_list,
                        iter_malichannel_list,
                        iter_freq_type_list,
                        iter_freqparavalue_list,
                        iter_biastype_list,
                        iter_biasparavalue_list
                      ]
```

### 4.1.3 Generation of Bias Vectors

This platform presents a sturdy framework that allows the generation and manipulation of attacks, facilitating examination of their impacts under diverse configurations and conditions. When an iteration attack case is fed into our bias generator, it results in four bias matrices corresponding to four vulnerable channels. Each matrix is structured with columns equivalent to the vehicle count and rows equivalent to the maximum possible iterations within a given control step. For instance, assuming $fv1$ and $fv3$ are victims and we have knowledge of all related parameters and the control step $k$ at which they are attacked, our algorithm would initially compute the bias for $fv1$ for all iteration steps at the control step $k$, and insert them into the first column of the bias vector. Since $fv2$ is not attacked, the second column will be filled with zeros. In a similar vein, the third column would contain some values, while all other columns would be zeros. Four bias vectors will be produced for the four vulnerable channels. The design is scalable, hence accommodating any new or zero-day attacks with relative ease. The algorithm employed to generate such bias vectors is demonstrated in object 4-1.

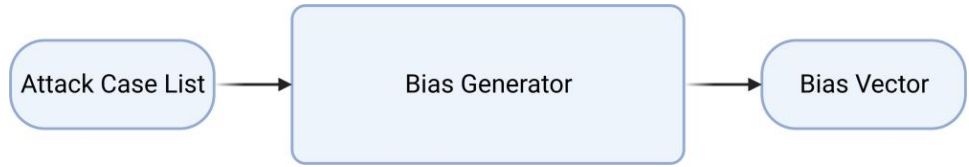

Figure 4-2. Bias Generator

### 4.1.4 Visualization of Bias Vectors

As depicted in Figure 4-2, the bias generator is responsible for producing bias vectors which are subsequently incorporated into malicious channels. This process manipulates the data exchanged in V2V communication. To accurately evaluate the bias generator's functionality, we have navigated through the extensive attack space in a systematic and methodical manner. The bias values have been carefully chosen to emulate the impacts that could occur in real-life situations. During V2V communication, these bias vectors are added to their corresponding channels to instigate an attack. For a comprehensive understanding and assessment of the output, we have included several illustrative figures. These visual aids encapsulate a broad spectrum of possible attack biases that could be generated by the system.

1: **function** ITERCHANNELBIAS(m, freqtype, freqparavalue, biastype, biasparavalue, maximum_iteration)
2:   Initialize stealth_mask and ite_bias_vec with zeros
3:   Configure stealth_mask based on freqtype[m]
4:   Set up ite_bias_vec based on biastype[m]
5:   **return** ite_bias_vec
6: **end function**
7: **function** ITERATTACKVALUE_CAL(n, k, maximum_iteration, iter_attackcase)
8:   Initialize x_ite_bias_df, v_ite_bias_df, zv_ite_bias_df, zx_ite_bias_df with zeros
9:   Extract parameters from iter_attackcase
10:  **for** each i in range of iter_victim_list **do**
11:    **for** each j in range of control_attackperiod_list[i] **do**
12:      Check condition on k and control_attackperiod_list[i][j]
13:      Extract values of malichannel, freqtype, freqparavalue, biastype, biasparavalue
14:      **for** each m in range of malichannel **do**
15:        Compute ite_bias_vec
16:        Update x_ite_bias_df, v_ite_bias_df, zv_ite_bias_df, zx_ite_bias_df based on malichannel[m]
17:      **end for**
18:    **end for**
19:  **end for**
20:  **return** x_ite_bias_df, v_ite_bias_df, zv_ite_bias_df, zx_ite_bias_df
21: **end function**

Object 4-1. Algorithm to Generate Bias Vector

Figure 4-3 *(a) Single Channel Single Target Bias* shows a bias vector where the 'x_ite' channel is compromised, and the target vehicle is *fv3* (Shown in green). The bias produced here is a 'Cluster' (On: Off period ratio is 1: 3), 'Linear' bias with a slope value of 0.2 and a constant value of 5 *m*. All other vehicles that are not targets have zero bias, as shown in the figure. Any bias is only generated if the control period *k* falls within the attack period list of the attack case. Otherwise, the output bias vector will have zeroes.

Figure 4-3 *(b) Single Channel Multi Target Bias* shows a bias vector where the 'v_ite' channel is compromised and the target vehicles are *fv2* (Shown in orange), *fv3* (Shown in green), and *fv4* (Shown in red). The bias generated for *fv2* is a 'Cluster' (On: Off period ratio is 1: 5), 'Sinusoidal' bias with an amplitude value of 20 $ms^{-1}$, frequency of 5 *Hz* and a constant value of 2 $ms^{-1}$. The bias generated for *fv3* is a 'Continuous', 'Linear' bias with a slope value of 0.05 and a constant value of 5 $ms^{-1}$. Meanwhile, the bias generated for *fv4* is a 'Continuous', 'Constant' bias with a constant value of 4 $ms^{-1}$.

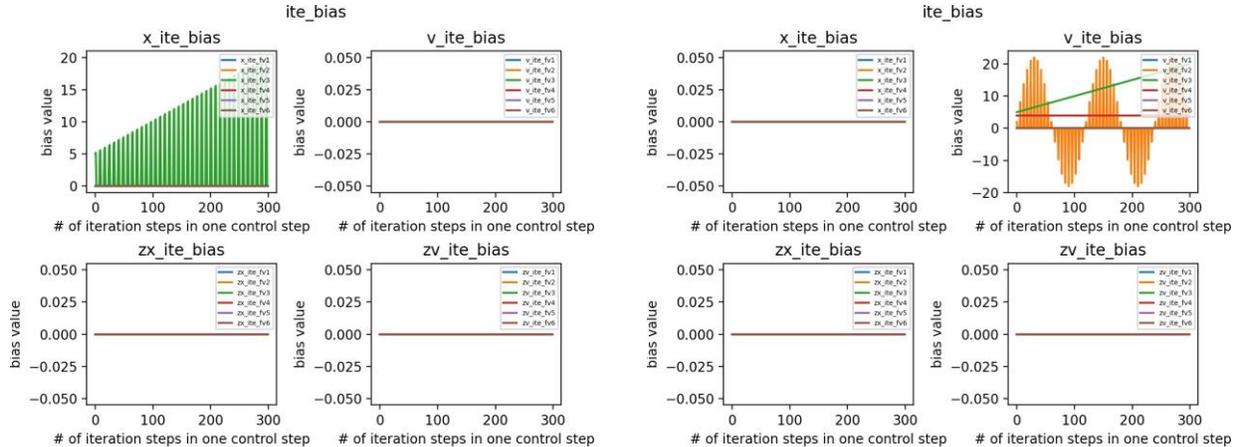

(a) Single Channel Single Target Bias    (b) Single Channel Multi Target Bias

Figure 4-3. Single Channel Attack Biases

Figure 4-4 *(a) All Channel Single Target Bias* shows a bias vector where all the 'x_ite', 'v_ite', 'zx_ite', and 'zv_ite' channel is compromised and the target vehicle is *fv4* (Shown in red). The bias generated for 'x_ite' is a 'Continuous', 'Sinusoidal' bias with an amplitude value of 15 *m*, frequency of 5 *Hz* and a constant value of 2 *m*. For 'v_ite' channel, the bias is a 'Continuous',

'Linear' bias with a slope value of $-0.05$ and a constant value of 5 $ms^{-1}$. For the channel 'zx_ite', the bias is 'Cluster' (On: Off period ratio is 2: 5), 'Sinusoidal' bias with an amplitude value of 20 $m$, frequency of 5 $Hz$ and a constant value of 2 $m$. Finally for the channel 'zv_ite', the bias is 'Cluster' (On:Off period ratio is 1:5), 'Constant' bias with a constant value of $-4$ $ms^{-1}$.

Figure 4-4 *(b) Multi Channel Single and Multi Target Bias* shows a complex bias vector where all the channels are compromised and the target vehicles are $fv2$ (Shown in orange), $fv3$ (Shown in green) and $fv4$ (Shown in red). The bias generated for $fv2$ is a 'Continuous', 'Sinusoidal' bias with an amplitude value of 20 $m$, frequency of 5 $Hz$ and a constant value of 2 $m$ for the 'x_ite' channel. Another compromised channel, 'zv_ite' for $fv2$ has a 'Cluster' (On: Off period ratio is 1: 5), 'Constant' bias with a constant value of 4 $ms^{-1}$. The bias generated on 'v_ite' channel for $fv3$ is a fusion of 'Continuous', 'Linear' and 'Continuous', 'Sinusoidal' bias with a slope value of 0.05 and a constant value of 5 $ms^{-1}$ for the linear function and an amplitude value of 20 $ms^{-1}$, frequency of 5 $Hz$ and a constant value of 2 $ms^{-1}$ for the sinusoidal function. Finally, the bias generated for $fv4$ on its 'zx_ite' channel is a 'Continuous', 'Constant' bias with a constant value of $-5$ $m$.

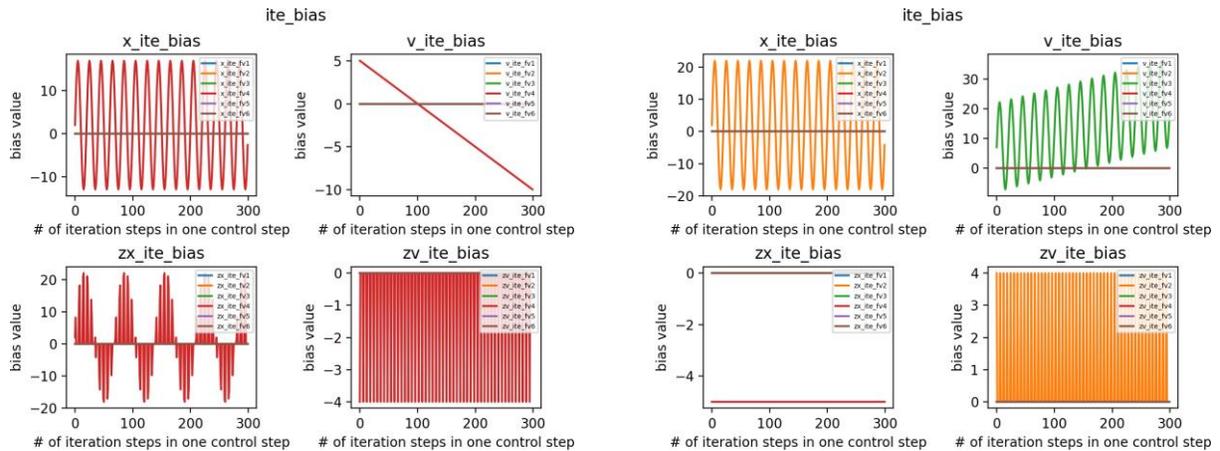

(a) All Channel Single Target Bias      (b) Multi Channel Multi Target Bias

Figure 4-4. Multi Channel Attack Biases

## 4.2  Impact Visualization and Analysis of V2V Perception Attacks

### 4.2.1  Concept of Attack and System Vulnerability

The utilization of attack impact visualization is a significant tool for security architects, providing them with a comprehensive view of resilience issues. This broader perspective is instrumental in understanding the complexities and vulnerabilities associated with the system. In the second chapter, we detailed a thorough categorization of possible attacks, which forms a crucial foundation for understanding how these attacks occur and the consequences they can produce.

In the context of V2V communications, an attack is typically executed during the iterative communication process among different vehicles. In our designed implementation, we induce an attack by injecting bias vectors into their corresponding communication channels. This manipulation of incoming data subsequently disrupts the iterative calculations of essential parameters such as ego position and velocity, as well as their relative position and velocity.

As a consequence of this disruption, the Dual loop conditions necessary for the optimal functioning of the system often fail to be satisfied. This failure leads to the premature termination of the loop once it reaches its maximum allowable number of iterations. The result of this premature termination is the inaccurate calculation of ego acceleration.

To illustrate this concept more clearly, consider the scenario where an attack bias is introduced to a target vehicle, say $fv3$. If the corrupted channels are 'x_ite' and 'v_ite', the information flow would lead to an impact on $fv4$. Similarly, if the corrupted channels are 'zx_ite' and 'zv_ite', the vehicle $fv2$ would bear the brunt of the attack according to the information flow.

### 4.2.2  Representative Attack Categories and Their Description

In this section, we explored various examples of representative perception attacks on the vehicle platoon, highlighting their impacts on the targeted vehicles. The primary indicators of this impact will be the resulting time-headway between the vehicles and the change in acceleration resulting from the attack since acceleration is the output of the MPC algorithm.

We provided visual representations, including graphs, to better illustrate safety and efficiency degradation attacks, along with the subsequent impacts that arise due to the application of the generated bias. It's important to note that the safe time headway is determined relative to speed. In our implementation, we assume an initial speed of 30 $ms^{-1}$ and thus consider a time-headway of $0.45s$ to $0.55s$ to be safe. Time-headway values below the lower threshold are deemed unsafe, while values above the upper threshold are viewed as contributing to efficiency degradation. All the studied attack categories are summarized in Table 4-1 for easy reference and understanding.

Table 4-1. Attack Category Description

| Attack Category | Frequency Type | Bias Type | Attack Channels | Attack Impact |
|---|---|---|---|---|
| 1 | Continuous | Constant | | |
| 2 | Continuous | Linear | x_ite | Efficiency |
| 3 | Continuous | Sinusoidal | v_ite | And |
| 4 | Cluster | Constant | zx_ite | Safety |
| 5 | Cluster | Linear | zv_ite | Degradation |
| 6 | Cluster | Sinusoidal | | |

Two representative attacks are shown in figure 4-5, which portrays how bias selection can result in safety and efficiency degradation. The control attack period for both attacks is $[40, 65]$. On the top row of the figures, the comparison of time-headway under the benign scenario (left) and time-headway under the attack scenario (right) can be observed. For better understanding of what happens to the acceleration, the comparison of acceleration under the benign scenario (left) and acceleration under the attack scenario (right) can be seen on the bottom row. Figure 4-5 is generated through a 'constant', 'continuous' attack bias on the 'x_ite' channel and with the target vehicle as $fv4$. However, from figure 4-5 (a) it can be observed that a positive bias of 10 $m$ had been applied which resulted in the fall of time-headway for $fv5$ below the safety level, thus making it a safety degradation attack. And, from figure 4-5 (b) it can be observed that a negative bias of 10 $m$ had been applied which resulted in the increase of time-headway way for $fv5$ above the safety level thus making it an efficiency degradation attack. So, from this example plot a clear idea can be obtained that the polarity of bias is the key deciding factor for an attack to compromise safety or efficiency. From both figures 4-5 (a) and (b) we can see that the vehicle undergoes rapid acceleration increase/decrease, which is alarming, and they have a significant deviation from the benign acceleration range $[-1.5, 1]$ $ms^{-2}$.

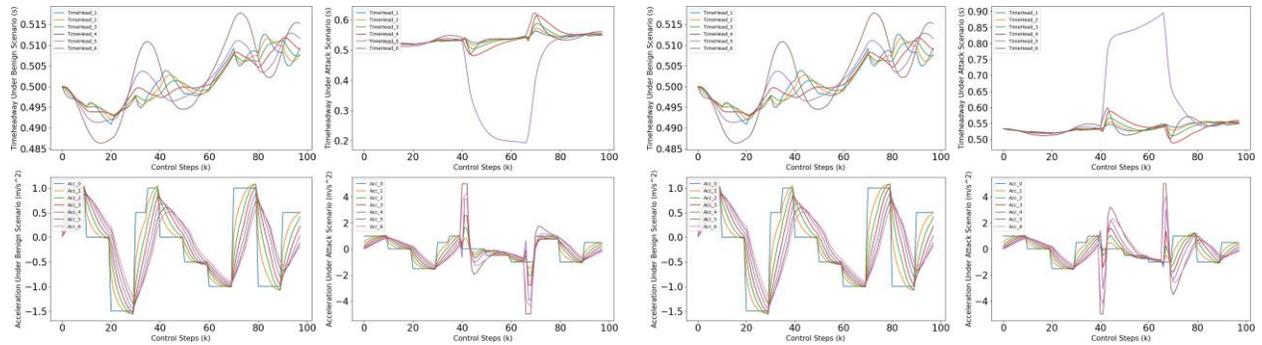

(a) Single Channel Single Target Safety Degradation Attack  (b) Single Channel Single Target Efficiency Degradation Attack

Figure 4-5. Safety and Efficiency Degradation Attack Impacts

Figure 4-6 (a) shows the attack impacts obtained by using the bias vector shown in figure 4-3 (a). The attack control period is selected to be $[40, 60]$. Upon observing the impact we can see it turns out to be a safety degradation attack. Similarly, figure 4-6 (b) shows the attack impacts obtained by using the bias vector shown in figure 4-3 (b). Here, as per implementation vehicles $fv3$, $fv4$ and $fv5$ are primarily affected when biases are added to vehicles $fv2$, $fv3$, $fv4$ during period $[20, 30]$, $[40, 50]$, $[70, 80]$ respectively. A combination of both safety and efficiency degradation can be observed and this happens when 'Sinusoidal' bias is given. This is also known as string instability.

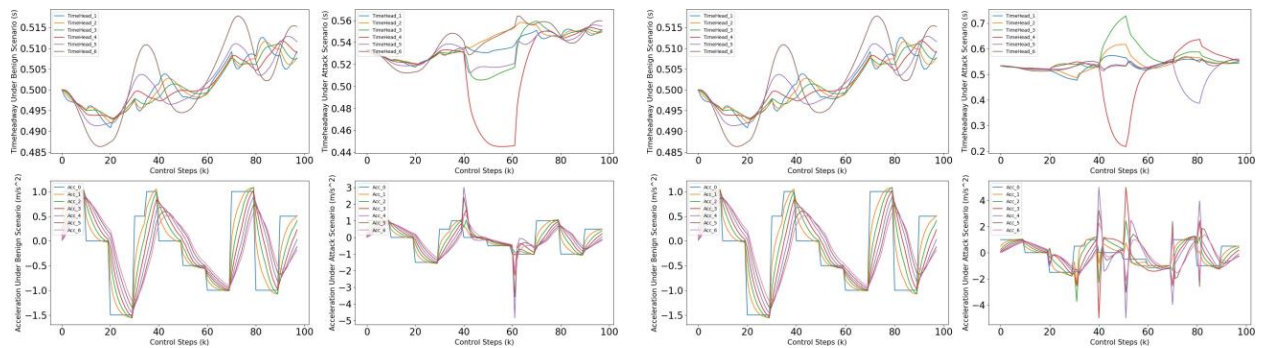

(a) Single Channel Single Target Attack  (b) Single Channel Multi Target Attack

Figure 4-6. Single Channel Attack Impacts

Likewise, figure 4-7 (a) shows the attack impacts obtained by using the bias vector shown in figure 4-4 (a) having attack control period selected to be $[20, 30]$. Upon observing the impact, we can see it turns out to be a efficiency degradation attack despite having positive and negative biases and sinusoidal biases. The season is all the biases acting upon the same vehicle creates a resultant impact. Similarly, figure 4-7 (b) shows the attack impacts obtained by using the bias vector shown in figure 4-4 (b). Here, as per implementation, vehicles $fv3$, $fv4$, and $fv5$ are primarily affected when biases are added to vehicles $fv2$, $fv3$, $fv4$ during period $[20, 30]$, $[45, 60]$, $[75, 85]$ respectively. The cumulative attack impacts on multiple vehicles in the platoon can be observed from the figure resulting from a very complex bias vector.

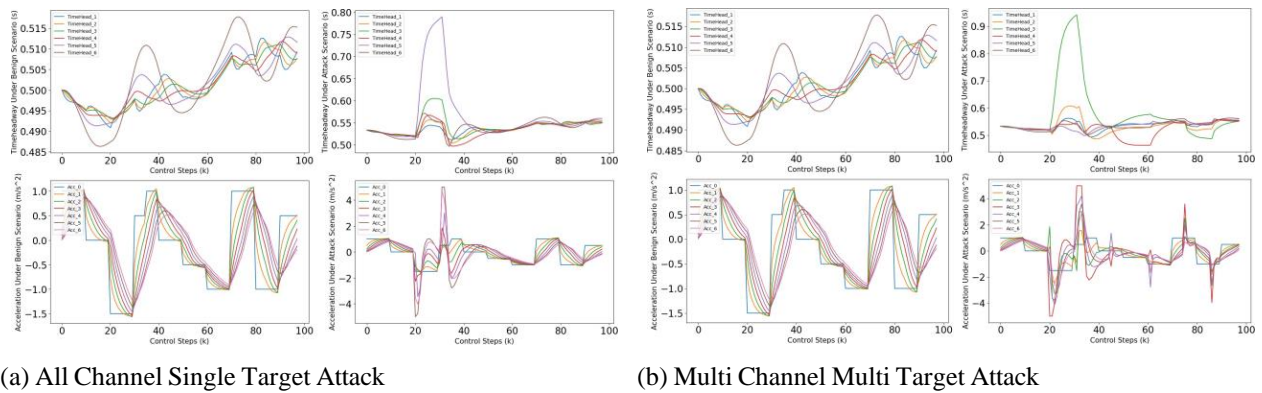

(a) All Channel Single Target Attack          (b) Multi Channel Multi Target Attack

Figure 4-7. Multi Channel Attack Impacts

We have also conducted investigations into the possibility of confining attacks within a single control step. However, to make such attacks effective, the bias introduced would have to be significantly high. Unfortunately, the high bias undermines the stealthiness of the attack, making it easily detectable and thus impractical in real-world scenarios. This lack of subtlety goes against one of the primary objectives of effective cyber-attacks, which is to remain undetected for as long as possible to maximize damage. Therefore, after careful consideration, we decided to exclude this approach from our overall attack strategy.

# CHAPTER 5
# DETECTION AND EVALUATION

After knowing the nature and tactics of the attacks, the next crucial element in reinforcing our system against such incursions is the ability to detect such attacks when they occur. It is critical that we have procedures in place that can detect and warn us of any anomalies that may indicate an ongoing attack.

This critical detection method is mostly based on detecting anomalies in the data. These deviations or abnormalities are often highlighted during the iteration phase, a critical aspect of the process that repeats over a range of control periods. During this iteration phase, data flows through four unique channels, giving us comprehensive coverage and the opportunity to record any unexpected patterns or values. It is critical to keep an eye on these channels since abnormal data can have far-reaching consequences for the system's safety and efficiency. These anomalies can have a direct impact on the 'ego vehicle,' which is our primary, and any change in the data could result in an undesired movement in the position or velocity of the ego vehicle. The immediate consequence would be unpredictability in the calculated acceleration.

To streamline the process and enhance clarity, we have designed a detection architecture. This structure, as depicted in the figure 5-1, presents an organized and detailed overview of how the detection mechanism functions, providing an illustrative representation of how the system works to identify and alert us of any potential attacks.

## 5.1 Detection Process and Limitations

### 5.1.1 Initial and Secondary Stages of Detection

A sophisticated dual-stage detection mechanism has been designed with the primary aim of determining if a vehicle is under attack or not. The process is as follows:

The initial detection stage is centered around monitoring any deviation in the position of the self-driving vehicle, often referred to as the 'ego vehicle'. This is achieved by calculating the difference between the distances of the ego vehicle to the one preceding it and the one following it. In circumstances where these gaps remain steady and their difference is minuscule, the system operates under the presumption that the vehicle is not under any form of attack. However, should these gaps show inconsistency and the difference breaches a predefined threshold, the system then interprets this as an indication of an attack. This methodology proves successful in identifying the majority of attacks.

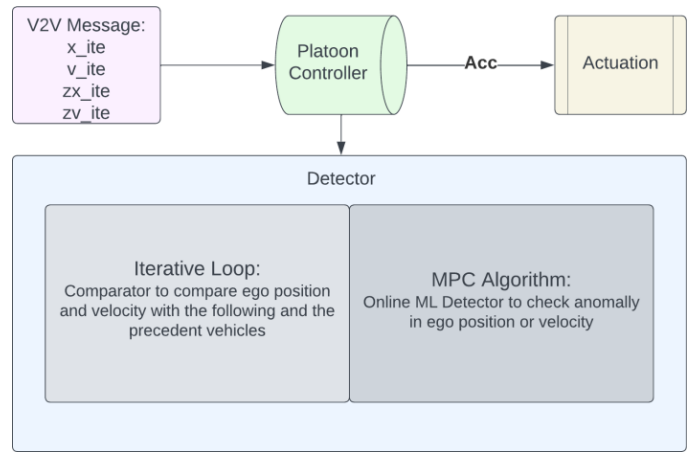

Figure 5-1. Corruption Detection Architecture

However, it is not without its limitations. For instance, if an attacker manipulates both the preceding and the following vehicles such that the distances (or gaps) increase or decrease uniformly, resulting in a difference that falls below the detection threshold, this method could be fooled into a false sense of security, thereby failing to detect the attack.

To counteract this potential vulnerability, a second stage of detection has been integrated, designed to strengthen the overall system. This reinforcement employs an online time series machine learning (ML) model embedded within the MPC algorithm, which is tasked with predicting the future position and velocity of the ego vehicle. If a significant deviation is found between the current calculated value and the predicted future value, the system flags this discrepancy as a potential attack.

**5.1.2 Understanding the Machine Learning Detector**

The machine learning model chosen for this task is the Extreme Learning Machines (ELM), a type of feed-forward neural network known for its impressive performance and rapid learning capabilities. The primary motivation for selecting this model lies in its ability to handle real-time problems effectively, courtesy of its swift online computation abilities. To prevent any contamination of the model's integrity due to the potential of training with corrupted data during an attack, a safeguard mechanism has been implemented. As soon as an attack is detected, this mechanism triggers the ELM model to stop its learning process and rely instead on previously trained weights. This measure ensures that the ML model is not inadvertently trained with tainted data, maintaining the robustness of the system. The algorithm for ELM model is shown in object 5-1.

**Require:** $L$ (number of hidden neurons), *random_state* (for reproducible results), *lag* (size of sliding window), *step_forward* (how far ahead to predict), and *series* (input data)

1: Define class *MinMaxNormalization* with new desired range.
2: **Within MinMaxNormalization:**
3:    Define method *fit* to find min and max values of data.
4:    Define method *transform* to scale data into new range.
5:    Define method *inverse_transform* to convert data back to original range.
6: Define class *ELMRegressor* with $L$ and *random_state*.
7: **Within ELMRegressor:**
8:    Define method *fit* to train the model.
9:    Define method *predict* to make predictions.
10: Define function *sliding_window* to prepare data for *ELMRegressor*.
11: **Within sliding_window:**
12:    Create input arrays (X) and output arrays (y) based on input data and *lag*.
13: Normalize *series* with *MinMaxNormalization*.
14: Prepare data for training with *sliding_window*.
15: **If** attack is not detected **then**
16:    Fit *ELMRegressor* with prepared data and save the model.
17: **Else**
18:    Load existing model.
19: Prepare input for prediction (last two values from normalized *series*).
20: Make prediction with *ELMRegressor*.
21: Convert prediction back to original range with *MinMaxNormalization inverse_transform*.
22: **return** Prediction

Object 5-1. ELM-based Anomaly Detection

## 5.2    Detection Analysis and Evaluation

In order to assess the effectiveness of the detection mechanism, we need to meticulously scrutinize the outputs of both the comparator and the machine learning detector. These two elements form the cornerstone of our detection mechanism, and a thorough understanding of their functioning is key to a comprehensive evaluation. We have attempted numerous test cases but have chosen to illustrate this evaluation process through two representative detection scenarios. These scenarios serve as practical examples which allow us to examine the detection mechanism in action and assess its performance in diverse situations. By observing how the comparator and the ML detector function under these different conditions, we can gain valuable insights into the strengths and weaknesses of our detection mechanism and, consequently, determine its overall efficacy.

### 5.2.1    Scenario 1: Single Target Attack with Anomaly Results

Let's first delve into the initial scenario. In this case, an attack is carried out on the controller using a bias vector to instigate a visible effect, as demonstrated in figure 4-5 (a). However, to facilitate efficient evaluation, a small period ($[40, 45]$) is utilized. Remarkably, the comparator was able to identify this intrusion independently, without any assistance from ELM. The output read "Anomaly Detected from control step 41" to "Anomaly Detected from control step 54". This detection lasted for an additional 9 control steps even after the cessation of the attack. The reason is that the controller requires a certain amount of time to return to its regular pattern via self-resilience. When we scrutinized the machine learning detector, it also exhibited a successful identification of the attack, as depicted in figure 5-2.

Moreover, table 5-1 provides a comprehensive summary of the specifics of the anomaly in question, including the type of anomaly, control steps, the primarily affected vehicle, the actual calculated values, and the predicted reference values for the aforementioned attack scenario.

### 5.2.2    Scenario 2: Multiple Target Attack with Anomaly Results

In the second instance, an assault is enacted on the controller using a bias vector to produce a significant effect, as shown in figure 4-7 (b). This case involves a more complex, fusion bias and thereby presents additional challenges. For this attack, the incursion was conducted during periods $[55, 60]$, $[70, 75]$, $[80, 85]$ for $fv3$, $fv4$, $fv5$, respectively.

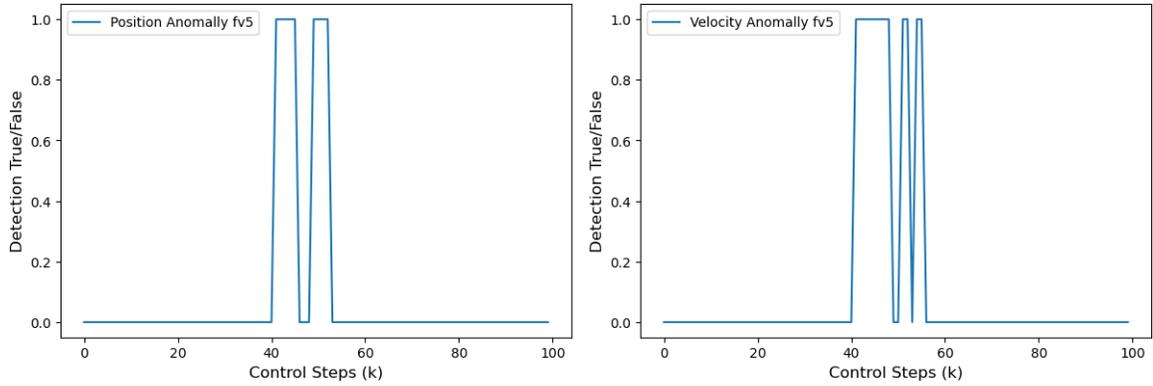

Figure 5-2. Anomaly Detection by ELM for Single Target

In this attack scenario, the comparator was able to detect the commencement of the attack from control step 56, but its detection ended at control step 76. Notably, there was an attack occurring from control step 80 to 85 that the comparator did not detect. This highlights the importance of the ML detector's role. As evident from figure 5-3, the ML detector successfully identified the attack even after control step 80.

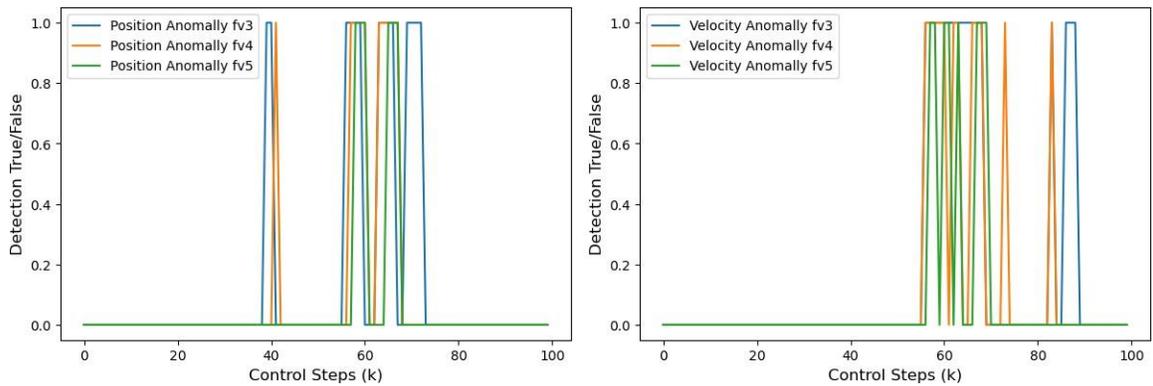

Figure 5-3. Anomaly Detection by ELM for Multiple Targets

Table 5-2 provides a detailed overview of the attributes of the detected anomaly, such as the type of anomaly, control steps, the primarily impacted vehicles, the actual calculated values, and the predicted reference values based on the above-mentioned attack scenario. It's worth noting, however, that there were a few instances of false detections during control steps 39 to 41, which is

Table 5-1. ML Detection For a Single Target Vehicle

| Anomaly Type | Control Step | Vehicle No. | Actual Value | Predicted Value |
|---|---|---|---|---|
| Pos Anom | 41 | 5 | 436.026 | 432.419 |
| Vel Anom | 41 | 5 | 32.994 | 29.602 |
| Pos Anom | 42 | 5 | 449.623 | 444.629 |
| Vel Anom | 42 | 5 | 34.994 | 29.952 |
| Pos Anom | 43 | 5 | 463.639 | 457.721 |
| Vel Anom | 43 | 5 | 35.086 | 37.369 |
| Pos Anom | 44 | 5 | 477.540 | 472.295 |
| Vel Anom | 44 | 5 | 34.417 | 39.290 |
| Pos Anom | 45 | 5 | 491.150 | 487.668 |
| Vel Anom | 45 | 5 | 33.634 | 41.293 |
| Vel Anom | 46 | 5 | 31.634 | 34.948 |
| Vel Anom | 47 | 5 | 29.634 | 31.950 |
| Vel Anom | 48 | 5 | 27.634 | 30.957 |
| Pos Anom | 49 | 5 | 538.643 | 542.500 |
| Pos Anom | 50 | 5 | 549.131 | 553.575 |
| Pos Anom | 51 | 5 | 559.845 | 563.832 |
| Vel Anom | 51 | 5 | 27.154 | 22.812 |
| Pos Anom | 52 | 5 | 570.856 | 573.509 |
| Vel Anom | 52 | 5 | 27.899 | 22.734 |
| Vel Anom | 54 | 5 | 29.123 | 31.612 |
| Vel Anom | 55 | 5 | 29.568 | 31.991 |

a known limitation inherent to any ML models.

Table 5-2. ML Detection For Multiple Target Vehicles

| Anomaly Type | Control Step | Vehicle No. | Actual Value | Predicted Value |
|---|---|---|---|---|
| Pos Anom | 39 | 3 | 448.134 | 445.532 |
| Pos Anom | 40 | 3 | 459.967 | 457.242 |
| Pos Anom | 41 | 4 | 453.188 | 450.634 |
| Pos Anom | 56 | 3 | 655.432 | 658.073 |
| Vel Anom | 56 | 3 | 26.744 | 30.480 |
| Vel Anom | 56 | 4 | 28.565 | 30.736 |
| Pos Anom | 57 | 3 | 665.805 | 670.161 |
| Pos Anom | 57 | 4 | 647.993 | 650.918 |
| Vel Anom | 57 | 3 | 25.120 | 30.240 |
| Vel Anom | 57 | 4 | 26.885 | 30.499 |
| Vel Anom | 57 | 5 | 28.312 | 30.696 |
| Pos Anom | 58 | 3 | 675.665 | 680.751 |
| Pos Anom | 58 | 4 | 658.466 | 662.445 |
| Pos Anom | 58 | 5 | 640.615 | 643.589 |
| Vel Anom | 58 | 3 | 24.178 | 27.428 |
| Vel Anom | 58 | 4 | 25.478 | 28.210 |
| Vel Anom | 58 | 5 | 27.000 | 29.986 |

Table 5-2. ML Detection For Multiple Target Vehicles (Continued)

| Anomaly Type | Control Step | Vehicle No. | Actual Value | Predicted Value |
|---|---|---|---|---|
| Pos Anom | 59 | 3 | 685.341 | 689.452 |
| Pos Anom | 59 | 4 | 668.519 | 672.694 |
| Pos Anom | 59 | 5 | 651.212 | 654.804 |
| Vel Anom | 59 | 3 | 24.202 | 29.560 |
| Vel Anom | 59 | 4 | 24.787 | 27.547 |
| Pos Anom | 60 | 4 | 678.407 | 681.784 |
| Pos Anom | 60 | 5 | 661.503 | 664.860 |
| Vel Anom | 60 | 3 | 24.639 | 31.520 |
| Vel Anom | 60 | 4 | 24.653 | 29.228 |
| Vel Anom | 60 | 5 | 25.467 | 27.777 |
| Vel Anom | 61 | 3 | 26.639 | 24.436 |
| Vel Anom | 61 | 5 | 25.685 | 28.006 |
| Vel Anom | 62 | 3 | 28.639 | 26.192 |
| Vel Anom | 62 | 4 | 27.296 | 24.560 |
| Pos Anom | 63 | 3 | 728.275 | 723.721 |
| Pos Anom | 63 | 4 | 710.398 | 707.708 |
| Vel Anom | 63 | 3 | 30.639 | 27.478 |
| Vel Anom | 63 | 4 | 28.887 | 26.141 |
| Vel Anom | 63 | 5 | 27.803 | 25.679 |
| Pos Anom | 64 | 3 | 740.930 | 735.295 |
| Pos Anom | 64 | 4 | 722.300 | 718.178 |
| Vel Anom | 64 | 3 | 32.639 | 34.652 |
| Pos Anom | 65 | 3 | 754.066 | 748.604 |
| Pos Anom | 65 | 4 | 734.728 | 730.278 |
| Pos Anom | 65 | 5 | 716.369 | 712.954 |
| Vel Anom | 65 | 3 | 33.040 | 37.230 |
| Pos Anom | 66 | 3 | 767.128 | 762.671 |
| Pos Anom | 66 | 4 | 747.349 | 743.262 |
| Pos Anom | 66 | 5 | 728.550 | 725.027 |
| Vel Anom | 66 | 3 | 32.269 | 40.035 |
| Vel Anom | 66 | 4 | 31.587 | 35.900 |
| Pos Anom | 67 | 4 | 759.912 | 757.062 |
| Pos Anom | 67 | 5 | 740.831 | 737.980 |
| Vel Anom | 67 | 3 | 31.231 | 40.609 |
| Vel Anom | 67 | 4 | 31.232 | 38.137 |
| Vel Anom | 67 | 5 | 30.728 | 35.224 |
| Vel Anom | 68 | 3 | 30.119 | 33.663 |
| Vel Anom | 68 | 4 | 30.616 | 34.693 |
| Vel Anom | 68 | 5 | 30.481 | 34.069 |
| Pos Anom | 69 | 3 | 803.925 | 806.466 |
| Vel Anom | 69 | 5 | 30.030 | 32.037 |
| Pos Anom | 70 | 3 | 815.321 | 818.571 |

Table 5-2. ML Detection For Multiple Target Vehicles (Continued)

| Anomaly Type | Control Step | Vehicle No. | Actual Value | Predicted Value |
|---|---|---|---|---|
| Pos Anom | 71 | 3 | 826.388 | 829.513 |
| Pos Anom | 72 | 3 | 837.226 | 839.948 |
| Vel Anom | 73 | 4 | 27.886 | 30.581 |
| Vel Anom | 83 | 3 | 28.786 | 32.192 |
| Vel Anom | 83 | 4 | 28.243 | 30.303 |
| Vel Anom | 86 | 3 | 26.756 | 29.200 |
| Vel Anom | 87 | 3 | 26.354 | 28.923 |
| Vel Anom | 88 | 3 | 26.072 | 28.570 |

# CHAPTER 6
# CONCLUSION AND FUTURE WORK

## 6.1 Conclusion

This thesis delves thoroughly into the world of CAVs within a V2X system, with a particular emphasis on the essential challenges of attack exploration and detection. It put light on the enormous potential for these technologies to revolutionize transportation by improving safety, fuel efficiency, and road use, especially through platooning. However, alongside the benefits, it drew attention to the considerable security vulnerabilities that may pose significant threats to these systems.

The exploration of potential attacks was thoroughly conducted, utilizing a comprehensive threat model that outlined potential adversary behaviors and a variety of V2V attack techniques such as MITM and FDI. This rigorous exploration of threats served as a valuable basis for understanding the impacts of potential attack scenarios and emphasized the necessity of proactive threat analysis in building a resilient system.

In response to these potential threats, the thesis offered a novel two-tiered detection technique capable of detecting anomalies indicating prospective attacks. The first stage, focusing on the positional monitoring of the 'ego vehicle', effectively flagged most attacks. Recognizing its vulnerability to sophisticated attacks, a second stage was created. This incorporated an online time series machine learning model into the MPC algorithm, predicting future ego vehicle positions and velocity, thus enhancing the detection capabilities.

Through various detection scenarios, the robustness of this dual-stage detection mechanism was put to the test, revealing its strengths and limitations. The thesis underscored the importance of continuous assessment and enhancement of these models to keep pace with evolving threats.

## 6.2 Future Work

Future research directions that have emerged from this work are numerous and fascinating. The primary focus should be on establishing powerful mitigation methods to successfully respond to attacks detected by the detection system. While the current work establishes a solid framework for detection, the next critical step is to ensure that these detections result in efficient and timely responses that protect the system from potential harm.

This system's resilience could be improved by utilizing data from trusted sensors. A resilient solution might alternatively be constructed using artificial intelligence or machine learning. Vehicle kinematics could be employed to achieve resiliency as well. Increasing the MPC controller's robustness by introducing additional constraints can also result in resiliency.

Furthermore, improving the dual-stage detection process by incorporating more powerful machine learning models and algorithms could be a worthwhile endeavor. Improving prediction and detection capabilities may result in a more proactive defense approach capable of repelling even more complicated attacks.

Finally, there is room to go deeper into the attacking spectrum. As technology improves, new types of threats will definitely emerge, necessitating the need to constantly update and refine the threat model.

Comprehensive real-world testing and validation of the decentralized platoon controller and the detection mechanisms would also be beneficial, enabling for more precise optimization and adaptation for practical applications. This thesis, therefore, serves as a stepping stone for future research aimed at securing resilient autonomous vehicle systems against evolving threats.

LIST OF REFERENCES

[1] F. Alotibi and M. Abdelhakim, *Anomaly detection for cooperative adaptive cruise control in autonomous vehicles using statistical learning and kinematic model*, IEEE Transactions on Intelligent Transportation Systems (2020), 1–11.

[2] Zoleikha Abdollahi Biron, Satadru Dey, and Pierluigi Pisu, *Real-time detection and estimation of denial of service attack in connected vehicle systems*, IEEE Transactions on Intelligent Transportation Systems **19** (2018), no. 12, 3893–3902.

[3] Roghieh A Biroon, Zoleikha Abdollahi Biron, and Pierluigi Pisu, *False data injection attack in a platoon of cacc: real-time detection and isolation with a pde approach*, IEEE Transactions on Intelligent Transportation Systems (2021).

[4] Srivalli Boddupalli, Ashwini Hegde, and Sandip Ray, *RePlaCe: Real-time security assurance in vehicular platoons against v2v attacks*, IEEE International Conference on Intelligent Transportation (2021).

[5] Srivalli Boddupalli, Richard Owoputi, Chengwei Duan, Tashfique Choudhury, and Sandip Ray, *Resiliency in connected vehicle applications: Challenges and approaches for security validation*, Proceedings of the Great Lakes Symposium on VLSI 2022 (New York, NY, USA), GLSVLSI '22, Association for Computing Machinery, 2022, p. 475–480.

[6] ______, *Machine learning for security resiliency in connected vehicle applications*, pp. 485–505, Springer International Publishing, Cham, 2023.

[7] Srivalli Boddupalli, Akash Someshwar Rao, and Sandip Ray, *Resilient cooperative adaptive cruise control for autonomous vehicles using machine learning*, IEEE Transactions on Intelligent Transportation Systems (2022).

[8] Joe Diether Cabelin, Paul Vincent Alpano, and Jhoanna Rhodette Pedrasa, *Svm-based detection of false data injection in intelligent transportation system*, 2021 International Conference on Information Networking (ICOIN), IEEE, 2021, pp. 279–284.

[9] Raj Gautam Dutta, Yaodan Hu, Feng Yu, Teng Zhang, and Yier Jin, *Design and analysis of secure distributed estimator for vehicular platooning in adversarial environment*, IEEE Transactions on Intelligent Transportation Systems (2020).

[10] S. Gong, J. Shen, and L. Du, *Constrained optimization and distributed computation based car-following control of a connected and autonomous vehicle platoon*, Transportation Research Part B: Methodological **94** (2016), 314–334.

[11] Xianhua Huang and Xinyu Wang, *Detection and isolation of false data injection attack in intelligent transportation system via robust state observer*, Processes **10** (2022), no. 7.

[12] Niloofar Jahanshahi and Riccardo MG Ferrari, *Attack detection and estimation in cooperative vehicles platoons: A sliding mode observer approach*, IFAC-PapersOnLine **51** (2018), no. 23, 212–217.

[13] Twan Keijzer and Riccardo MG Ferrari, *A sliding mode observer approach for attack detection and estimation in autonomous vehicle platoons using event triggered communication*, 2019 IEEE 58th Conference on Decision and Control (CDC), IEEE, 2019, pp. 5742–5747.

[14] Eshaan Khanapuri, Veera Venkata Tarun Kartik Chintalapati, Rajnikant Sharma, and Ryan Gerdes, *Learning based longitudinal vehicle platooning threat detection, identification and mitigation*, IEEE Transactions on Intelligent Vehicles (2021).

[15] Pengyuan Lu, Limin Zhang, B Brian Park, and Lu Feng, *Attack-resilient sensor fusion for cooperative adaptive cruise control*, 2018 21st International Conference on Intelligent Transportation Systems (ITSC), IEEE, 2018, pp. 3955–3960.

[16] Patrick M. Mendoza, Tashfique Hasnine Choudhury, and Sandip Ray, *Poster: Vehicle-to-infrastructure security for reduced speed work zone*, Proceedings of the Twenty-Fourth International Symposium on Theory, Algorithmic Foundations, and Protocol Design for Mobile Networks and Mobile Computing (New York, NY, USA), MobiHoc '23, Association for Computing Machinery, 2023, p. 571–573.

[17] Hassan Mokari, Elnaz Firouzmand, Iman Sharifi, and Ali Doustmohammadi, *Deception attack detection and resilient control in platoon of smart vehicles*, 2022 30th International Conference on Electrical Engineering (ICEE), IEEE, 2022, pp. 29–35.

[18] Alberto Petrillo, Antonio Pescape, and Stefania Santini, *A secure adaptive control for cooperative driving of autonomous connected vehicles in the presence of heterogeneous communication delays and cyberattacks*, IEEE transactions on cybernetics **51** (2020), no. 3, 1134–1149.

[19] Xiulan Song, Xiaoxin Lou, Junwei Zhu, and Defeng He, *Secure state estimation for motion monitoring of intelligent connected vehicle systems*, Sensors **20** (2020), no. 5, 1253.

[20] Michael Wolf, Alexander Willecke, Johannes-Christian Müller, Keno Garlichs, Thomas Griebel, Lars Wolf, Michael Buchholz, Klaus Dietmayer, Rens W van der Heijden, and Frank Kargl, *Securing cacc: Strategies for mitigating data injection attacks*, 2020 IEEE Vehicular Networking Conference (VNC), IEEE, 2020, pp. 1–7.

[21] LOU Xiaoxin, SONG Xiulan, HE Defeng, and MENG Liming, *Secure estimation for intelligent connected vehicle systems against sensor attacks*, 2019 Chinese Control Conference (CCC), IEEE, 2019, pp. 6658–6662.

[22] Xiaofei Zhang, Haiping Du, Jumei Wei, Zhijuan Jia, Suimin Jia, and Ge Ma, *High gain observer design for dos attack detection in cacc platoon*, 2020 International Seminar on Intelligent Technology and Its Applications (ISITIA), IEEE, 2020, pp. 254–259.

[23] Chunheng Zhao, Jasprit Singh Gill, Pierluigi Pisu, and Gurcan Comert, *Detection of false data injection attack in connected and automated vehicles via cloud-based sandboxing*, IEEE Transactions on Intelligent Transportation Systems (2021).

[24] Zepeng Zhou, Fanglai Zhu, Dezhi Xu, Shenghui Guo, and Younan Zhao, *Attack resilient control for vehicle platoon system with full states constraint under actuator faulty scenario*, Applied Mathematics and Computation **419** (2022), 126874.

[25] Zhiqiang Zuo, Xiong Cao, and Yijing Wang, *Security control of multi-agent systems under false data injection attacks*, Neurocomputing **404** (2020), 240–246.

BIOGRAPHICAL SKETCH

Tashfique Hasnine Choudhury is a skilled engineer who has a profound passion for using cutting-edge machine learning to further the development of autonomous and electric vehicles. He began his undergraduate studies at the Islamic University of Technology in Bangladesh, where he graduated in December 2019 with a bachelor of science in electrical and electronics engineering. He traveled to the United States to pursue his master's at the esteemed University of Florida because he was eager to increase his knowledge.

Tashfique, a specialist in the Department of Electrical and Computer Engineering, approached his research with great enthusiasm. His primary focus was on utilizing machine learning techniques to advance autonomous and electric vehicles, a pioneering field that has the potential to bring about a revolutionary change in transportation.

He worked tirelessly until August 2023, when he successfully earned his master of science degree. Tashfique has demonstrated through his challenging academic career that he is a committed researcher and an inventive engineer, working hard at the nexus of technology and sustainability.